\documentclass[twocolumn,preprintnumbers,amsmath,superscriptaddress,amssymb,prb]{revtex4-1}
\usepackage{graphicx}
\usepackage{dcolumn}
\usepackage{multirow}
\usepackage{bm}
\usepackage{booktabs}
\usepackage[colorlinks=true,linkcolor=blue, citecolor=blue, urlcolor=blue]{hyperref}
\usepackage{float}
\usepackage[normalem]{ulem}

\begin{document}

\title{Tuning Topological Phase Transitions in Hexagonal Photonic Lattices Made of Triangular Rods}
\author{Hsun-Chi Chan}
\address{Department of Physics and Center for Theoretical Physics, National Taiwan University, Taipei 10617, Taiwan}
\author{Guang-Yu Guo}
\email{gyguo@phys.ntu.edu.tw}
\address{Department of Physics and Center for Theoretical Physics, National Taiwan University, Taipei 10617, Taiwan}
\address{Physics Division, National Center for Theoretical Sciences, Hsinchu 30013, Taiwan}

\date{\today}

\begin{abstract}
In this paper, we study topological phases in a 2D photonic crystal with broken time ($\mathcal{T}$)
and parity ($\mathcal{P}$) symmetries by performing calculations of 
band structures, Berry curvatures, Chern numbers, edge states and also numerical simulations
of light propagation in the edge modes. Specifically, we consider a hexagonal lattice consisting of
triangular gyromagnetic rods. Here the gyromagnetic material breaks
$\mathcal{T}$ symmetry while the triangular rods breaks $\mathcal{P}$ symmetry.
Interestingly, we find that the crystal could host quantum anomalous Hall (QAH) phases 
with different gap Chern numbers ($C_g$) including $|C_g| > 1$
as well as quantum valley Hall (QVH) phases with 
contrasting valley Chern numbers ($C_v$), depending on the orientation of the triangular rods.
Furthermore, phase transitions among these topological phases, such as from QAH to QVH
and vice versa, can be engineered by a simple rotation of the rods.
Our band theoretical analyses reveal that the Dirac nodes at the $K$ and $K'$ valleys
in the momentum space are produced and protected by the mirror symmetry ($m_y$) instead
of the $\mathcal{P}$ symmetry, and they become gapped
when either $\mathcal{T}$ or $m_y$ symmetry is broken, resulting in a QAH or QVH
phase, respectively. Moreover, a high Chern number ($C_g = -2$) QAH phase
is generated by gapping triply degenerate nodal points rather than pairs
of Dirac points by breaking $\mathcal{T}$ symmetry.  
Our proposed photonic crystal thus provides a platform for investigating intriguing 
topological phenomena which may be challenging to realize in electronic systems,
and also has promising potentials for device applications in photonics such as 
reflection-free one-way waveguides and topological photonic circuits.
\end{abstract}


\maketitle

\section{Introduction}\label{sect1}

In recent years, electronic and photonic topological insulators\cite{kane10,joan14} 
have attracted enormous attention because these systems exhibit fascinating wave transport
properties. In particular, the gapless edge states on the surface or at the interface
between these topological insulators are unidirectional 
and robust against scattering from disorder due to topologically  
nontrivial properties of their bulk band structures. 
The electronic quantum anomalous Hall (QAH) phase, first proposed by Haldane\cite{hald88},
is a two-dimensional (2D) bulk ferrromagnetic insulator (Chern insulator) with a nonzero topological 
invariant called Chern number in the presence of spin-orbit coupling (SOC) but in the absence of 
applied magnetic fields.\cite{Weng15} Its associated chiral edge states  
carry dissipationless unidirectional electric current. 
Excitingly, this remarkable QAH phase was recently observed in ferromagnetic topological insulator films\cite{Chang13}. 
Moreover, Haldane and Raghu recently proposed the optical analogs of this intriguing QAH phase in 
photonic crystals made of time-reversal ($\mathcal{T}$) symmetry broken materials\cite{hald08a,hald08b}. 
This photonic QAH phase has gapless edge states within each topologically nontrivial bulk band gap,
and the number of the edge states is determined by the gap Chern number\cite{hald08b}. 
Such topologically protected edge states are immune to backscattering and are therefore robust
against disorder.\cite{joan08,joan09} Subsequently, the photonic topological phases in a number of 
gyromagnetic photonic crystals with broken $\mathcal{T}$ symmetry were  
proposed \cite{joan08,he10,wang13,skir14} and observed\cite{joan09,skir15}. 

When designing a photonic Chern insulator, one usually starts with a lattice with both $\mathcal{T}$ symmetry
and inversion ($\mathcal{P}$) symmetry where doubly degenerate Dirac points may exist
at some high symmetry points.\cite{hald08b} In particular, in a 2D lattice with both $\mathcal{P}$
and $\mathcal{T}$ symmetries, the stability of the Dirac points is guaranteed by the Wigner-von Neumann theorem.
Furthermore, one can find the double degeneracies by varying just one or two system parameters.
When $\mathcal{T}$ symmetry is broken, the Dirac points become gapped, resulting in a QAH phase.
Nevertheless, the topological phases in broken $\mathcal{P}$ symmetry Chern insulators 
have received much less attention, where different mechanisms would be needed.\cite{he15,Ono09} 

Furthermore, majority of the predicted or realized photonic Chern insulators so far have been limited to
the Chern number $|C|=1$,\cite{joan08} although in principle the Chern number can be any
integer values. Consequently, having systems with $|C| > 1$ is of fundamental interest in studying
topological phases. Systems with higher Chern numbers also have practical values. 
For example, they are useful for designing novel topological devices such as one-way photonic circuits.\cite{he10,wang13}  
Indeed, it has been recently proposed\cite{skir14} and demonstrated \cite{skir15} that
photonic Chern insulators of large Chern numbers can be realized in 2D square and hexagonal lattices
made of cylindrical rods by tuning the radius of the rods. It would be interesting to 
find other ways to realize photonic Chern insulators with $|C| > 1$.

Interestingly, recent progress in understanding the novel properties of electronic 2D materials such as graphene
and MoS$_2$ monolayer has led to the discovery of a $\mathcal{T}$ invariant topological phase called 
qunatum valley Hall (QVH) state in broken $\mathcal{P}$ symmetry materials\cite{Xiao07,Niu10,Mak14}.  
In these QVH materials, the energy extrema (or valleys) of the band structure 
at the $K$ and $K'$ points in their hexagonal Brillouin zone have contrasting properties such as
nonzero Berry curvatures of opposite signs due to the absence of $\mathcal{P}$ symmetry but presence
of $\mathcal{T}$ invariance, 
although they are energetically degenerate.\cite{Xiao07,Niu10,Mak14,Ma16,Chen17} 
This broken valley symmetry results in a number of interesting valley-contrasting phenomena
and also a totally new concept of electronics known as valleytronics\cite{Xiao07}.
Recently, this QVH effect has also been realized in all-dielectric photonic crystals\cite{Ma16,Chen17}
and bianisotropic metamaterials\cite{dong16}, and the valley-protected reflection-free 
propagation of light in the edge modes of these materials has been demonstrated. 
This would lead to the fascinating prospect of optical communication devices based on the robust flow of
light. 

In this paper, we explore possible topological phases in a 2D photonic crystal with broken $\mathcal{T}$ 
and $\mathcal{P}$ symmetries. We consider a hexagonal lattice made of 
triangular gyromagnetic rods, as illustrated in Fig. 1(a). Here the gyromagnetic material breaks
$\mathcal{T}$ symmetry while the triangular rods breaks $\mathcal{P}$ symmetry.
Interestingly, we find that the crystal hosts QAH phases with different gap Chern numbers including $|C| > 1$ 
and QVH phases with contrasting valley Chern numbers, depending on the orientation
[i.e., the rotation angle $\varphi$ in Fig. 1(b)] of the triangular rods. 
Furthermore, phase transitions among these topological phases, such as from QAH to QVH
and vice versa, can be engineered by a simple rotation of angle $\varphi$.
Our band theoretical analyses reveal that the Dirac points at the $K$ and $K'$ valleys
are produced and protected by the mirror symmetry ($m_y$) instead
of the $\mathcal{P}$ symmetry, and they become topologically gapped
when either $\mathcal{T}$ or $m_y$ symmetry is broken, resulting in a QAH or QVH
phase, respectively. Moreover, the high Chern number ($C_g = -2$) QAH phase
results from gapping triply degenerate nodal points rather than pairs
of Dirac points\cite{skir14} by breaking $\mathcal{T}$ symmetry.
Thus, our proposed photonic crystal offers a platform for investigating a number of
topological phenomena which may be challenging to realize in electronic systems,
and also has promising device applications in photonics.

The rest of this paper is organized as follows. First, we introduce the proposed structure 
and also describe the computational methods in the next section. 
Then the main results are presented in Sec. \ref{sect3}, including the rich topological gap map 
on the frequency-rod angle plane in Subsec.\ref{sect3a}, the representative 
bulk band structures in Subsec.\ref{sect3b}, the calculated Berry curvatures 
in Subsec. \ref{sect3c}, the gapless edge states in Subsec.\ref{sect3d} and reflection immunne
one-way waveguides in \ref{sect3e} as well as valley Hall edge states and light propagation in
Z-shape bends in \ref{sect3f}. Finally, the conclusions drawn from
this work are given in Sec. \ref{sect4}.

\begin{figure}
\includegraphics[width=7cm]{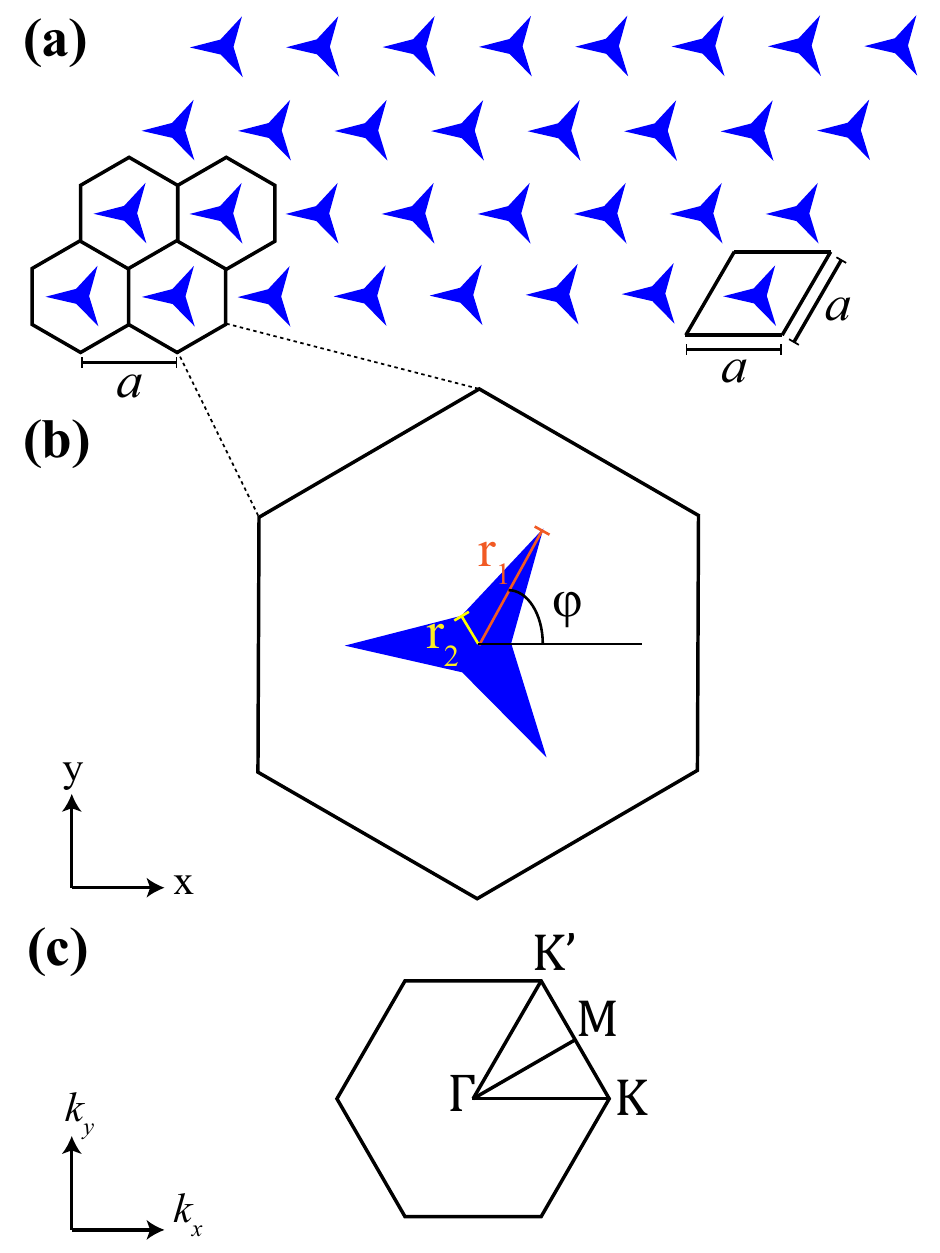}
\caption{\label{fig1}
(a) Schematic diagram of the hexagonal lattice of triangular gyromagnetic rods in air considered in the present paper.
Here $a$ is the lattice constant. (b) Enlarged unit cell containing one triangular rod of $\epsilon=13$, $\mu=1$, 
$\kappa=0.4$, $r_1=0.3075a$ and $r_2=0.0615a$. The first Brillouin zone of the hexagonal lattice
with  high symmetry points labelled.}

\end{figure}

\section{Structure and Computational method}\label{sect2} 
Here we consider a 2D photonic crystal consisting of triangular gyromagnetic rods arranged in a hexagonal 
lattice filled with air, as shown in Fig. \ref{fig1}(a). The $\mathcal{T}$ symmetry is broken by the
gyromagnetic rods used. The permeability tensor of the gyromagnetic material in SI units can be expressed as
\begin{equation} 
\label{eq4} 
\bm{\mu}=\mu_0
\begin{pmatrix}
\mu_r & i\kappa & 0 \\
-i\kappa & \mu_r & 0 \\
0 & 0 & 1
\end{pmatrix}.
\end{equation}
Following Ref. [\onlinecite{joan08}], we neglect the small loss and dispersion, and thus set the material parameters 
$\epsilon_r=13$, $\mu_r=1$ and $\kappa=0.4$. Note that these parameter values are close to such real gyromagnetic 
materials as yttrium-iron-garnet,\cite{skir15,joan09} which has $\mu_r=0.9$ and $\kappa=0.4$ at $13$ GHz 
with $0.1$ T static magnetic field.\cite{MWbook} 

The triangular rods are determined by two geometric parameters $r_1$ and $r_2$,
as indicated in Fig. \ref{fig1}(b). In our numerical simulations, we set $r_1=0.3075 a$ and $r_2=0.0615 a$ 
where $a$ is the lattice constant. Nevertheless, our conclusions will not be affacted by this perticaular 
choice of the parameters. The structure will be completely fixed only when the angle ($\varphi$) between 
one of the arms and the $x$-axis is specified [see Fig. \ref{fig1}(b)], and this angle is the principal 
parameter that one tunes to manipulate the properties of the structure, as will be presented in the next section. 
The structure has the $C_{3v}$ symmetry with 3 threefold $C_3$ rotations and 3 mirror ($m_v$) reflections, 
for $\varphi=n\times30^{\circ}$ and $n=0,1,2,3,4,5$. 
The mirror symmetries contain $m_y$ (reflection plane normal to $\hat{y}$) if $n=0,2,4$ but 
include $m_x$ (reflection plane normal to $\hat{x}$) if $n=1,3,5$.
The Brillouin zone (BZ) is also hexagonal, as shown in Fig. \ref{fig1}(c). However, the irreducbile
BZ wedge (IBZW) depends on angle $\varphi$. 
For example, when $\varphi=0^{\circ}$, the IBZW is $\Gamma$KM$\Gamma$ since K and K$'$ 
are equivalent. In contrast, when $\varphi=30^{\circ}$, the IBZW is $\Gamma$KMK$'\Gamma$
because K and K$'$ are no longer equivalent [see Fig. \ref{fig1}(c)].

To calculate the band structure of the proposed structure, we solve the Maxwell$'$s wave equation 
\begin{equation}
\label{eq1}
\nabla \times [\bm{\mu^{-1}(\mathbf{r})} \nabla \times \mathbf{E(r)}]=\bm{\epsilon(\mathbf{r})}\omega^2 \mathbf{E}
\end{equation} 
where $\bm{\mu}(\mathbf{r})$ and $\bm{\epsilon(\mathbf{r})}$ are the permeability and permittivity tensors, respectively, 
and $\omega$ is the eigenfrequency. We use the finite-element method implemented in the commercial software COMSOL
 Multiphysics \textregistered{}.\cite{comsol} 
To examine the topological nature of a band gap in the band structure, we also calculate the band Chern number
\begin{equation} 
\label{eq2} 
C_n=\frac{1}{2\pi}\int\limits_{BZ}\mathbf{\Omega}_n(\mathbf{k})\cdot d^2\mathbf{k},
\end{equation}
where the integral is over the BZ and $\mathbf{\Omega}_n(\mathbf{k})$ 
is the Berry curvature of the $n$th band defined as \cite{Niu10} 
\begin{equation} 
\label{eq3} 
\mathbf{\Omega}_n(\mathbf{k})=\mathbf{\nabla}_{\mathbf{k}} \times \mathcal{A}_n(\mathbf{k})
\end{equation}
where 
$\mathbf{\mathcal{A}}_n(\mathbf{k})=\langle E_{n\mathbf{k}} |i \mathbf{\nabla}_{\mathbf{k}} | E_{n\mathbf{k}} \rangle$
is the Berry connection and $E_{n\mathbf{k}}$ is the $n$th band energy. Here we adopt the efficient numerical 
algorithm reported in Ref. [\onlinecite{Hats05}] to calculate the band Chern numbers. 
Note that the Berry curvature $\mathbf{\Omega}_n$ and 
the Chern number $C_n$ are invariant under gauge transformation, 
while $\mathbf{\mathcal{A}}_n$ is gauge-dependent. 
Consequently, one advantage of the algorithm is that we can obtain $C_n$ via Eq. \eqref{eq2} 
without any gauge-fixing process.  Thanks to the efficiency of the algorithm,\cite{Hats05}  
we can obtain the accurate $C_n$ even using a moderate dense $k$-point
mesh, and this enables us to perform  massive calculations for searching candidate structures or 
plotting topological gap map such as Fig. 2.\cite{skir15} 

The integer Chern number $C_n$ is the topological invariant of the $n$th energy band. 
The sum of the Chern numbers of all the bands below a band gap is 
called the gap Chern number $C_{g}$.\cite{Hats93,skir14} 
According to the bulk-edge correspondence,\cite{Hats93} the number of gapless edge states 
between two topologically distinct materials equals to the gap Chern number difference 
across the interface.

\section{Results and discussion}\label{sect3}
\subsection{Topological gap map}\label{sect3a}
\begin{figure}
\includegraphics[width=8cm]{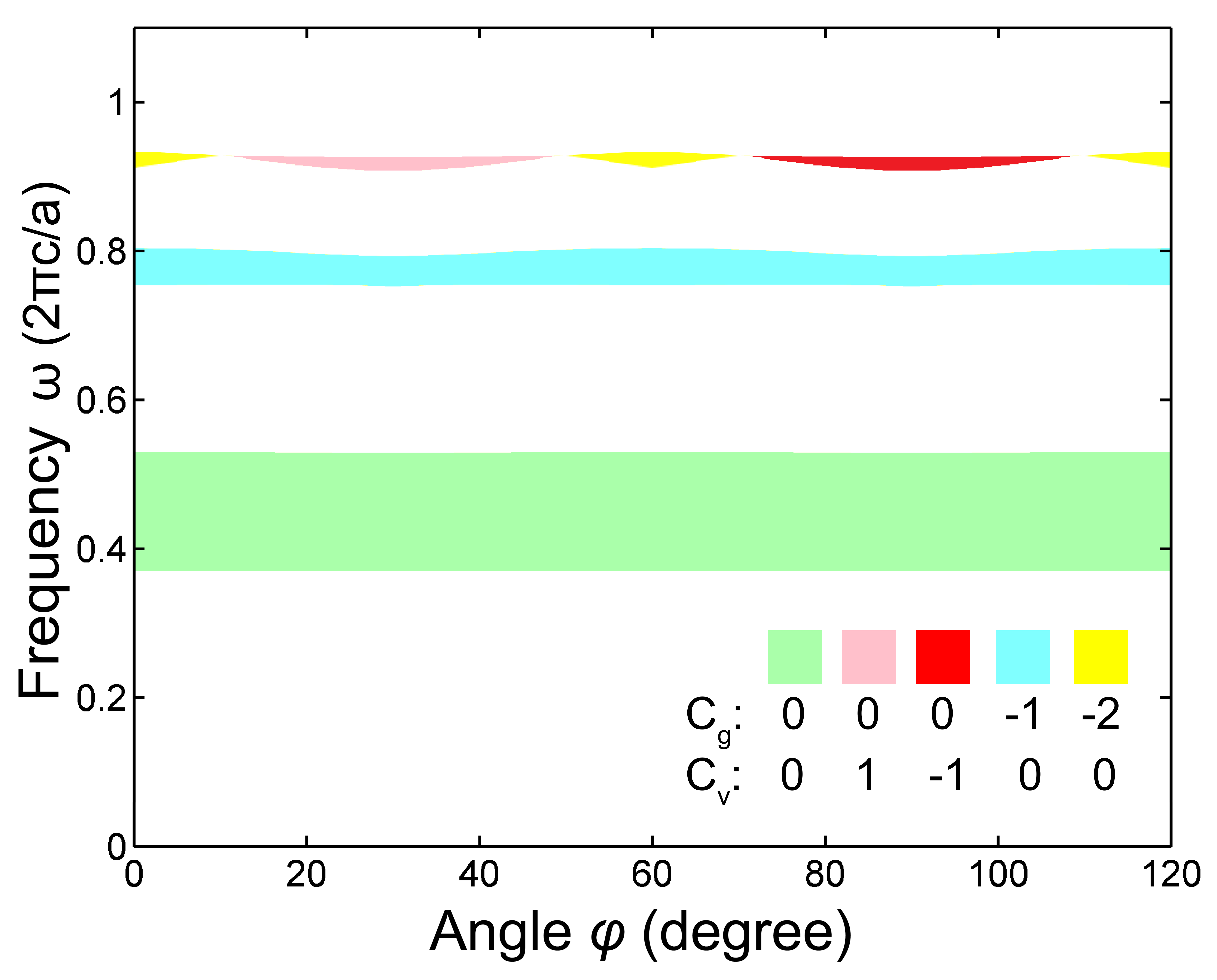}
\caption{\label{gapmap}
Topological band gap map on the plane of rotation angle $\varphi$ and frequency $\omega$.
Rotation angle $\varphi$ is defined in Fig. \ref{fig1}(b).
Each band gap region is labeled with a color representing its gap Chern number $C_g$ 
and valley Chern number $C_v$. Note that
in the third band gap region, as $\varphi$ increases, a series of topological phase transitions occur 
from $C_{g}=-2$ ($C_v =0$) to $C_{g}=0$ ($C_v =1$) at $\varphi = \sim 10^{\circ}$,
back to $C_{g}=-2$ ($C_v =0$) at $\varphi = \sim 50^{\circ}$, then to $C_{g}=0$ ($C_v =-1$)
at $\varphi = \sim 70^{\circ}$, and finally back to $C_{g}=-2$ ($C_v =0$) at $\varphi = \sim110^{\circ}$. }
\end{figure}

To have an overall picture of the topological phases and phase transitions in the proposed triangular photonic crystal, 
we construct a topological gap map in the plane of rotation angle $\varphi$ and frequency $\omega$,
as displayed in Fig. \ref{gapmap}.  
Since we find that the period of the rotation angle is $120^{\circ}$, we calculate the band gaps as 
a function of  the rotation angle from $\varphi=0^{\circ}$ 
to $\varphi=120^{\circ}$ with the angle step $\Delta \varphi = 1^{\circ}$. 

\begin{table}[h]
\caption{Calculated band gap ($\Delta \omega_g$), relative band gap ($\Delta \omega_g/\omega_m$),
gap Chern number ($C_g$), valley Chern number ($C_v$) and  number of edge states ($N_g$) for the 
$\varphi=0^{\circ}$ and $\varphi=30^{\circ}$ cases. Here $\Delta \omega_g=\omega_2-\omega_1$,
gap center $\omega_m=(\omega_1+\omega_2)/2$ where $\omega_1$ ($\omega_2$) is
the highest (lowest) frequency of the lower (upper) band of the gap. 
As a measure of unidirectionality, the edge state intensity ratio ($I_{-}/I_{+}$) 
of wave propagation along the interface between the crystal and the metal wall (see Subsec. \ref{sect3e}) 
is also listed. Here $I_{-}$ and $I_{+}$ are the averaged intensities at $x=-20a$ and $x=6a$, respectively.}
\begin{center}
\begin{tabular}{r|rrr|rrr} \hline\hline
structure     & \multicolumn{3}{c|}{$\varphi=0^{\circ}$} & \multicolumn{3}{c}{$\varphi=30^{\circ}$} \\
gap number    & $1^{st}$  & $2^{nd}$   & $3^{rd}$   & $1^{st}$  & $2^{nd}$   & $3^{rd}$   \\ \hline
$\Delta\omega_g$ ($2\pi c/a$)   & 0.16 & 0.056 & 0.021 & 0.16 & 0.041 & 0.018 \\
$\Delta\omega_g/\omega_m$ ($\%$) & 36 & 7.3 & 2.3 & 35 & 5.2 & 2.0 \\
$C_g$ & 0 & -1 & -2 & 0 & -1 & 0 \\
$C_v$ & 0 &  0 &  0 & 0 &  0 & 1 \\
$N_g$ & 0 & 1 & 2 & 0 & 1 & 0 \\
$I_{-}/I_{+}$ ($10^3$) & - & 1400 & 6 & - & 23 & - \\ \hline\hline
\end{tabular}
\end{center}
\end{table}

Figure 2 shows that there are three band gap regions in the frequency range from 0 to 1 $2\pi c/a$.
The topological nature of a band gap in a broken $\mathcal{T}$ symmetry system
can be characterized by the gap Chern number ($C_{g}$), which is the sum of the Chern numbers
of all the bands below the band gap, as mentioned above in Sec. II.
Therefore, the first band gap near 0.5 $2\pi c/a$ is topologically trivial because its gap Chern number
$C_{g}=0$. The second band gap just below 0.8 $2\pi c/a$ is topologically nontrivial 
and its gap Chern number $C_{g}=-1$. Moreover, the band gap is rather large, being around 
7 \% (0.056 $2\pi c/a$). Interestingly, when rotationa angle $\varphi$ varies from 0 to 120$^{\circ}$, 
the third band gap experiences a series of topology changes (see Fig. 2). 
For example, as $\varphi$ increases, the topological phase changes 
from $C_{g}=-2$ ($C_v =0$) to $C_{g}=0$ ($C_v =1$) at $\varphi = \sim10^{\circ}$,
and from $C_{g}=-2$ ($C_v =0$) to $C_{g}=0$ ($C_v =-1$) at $\varphi = \sim70^{\circ}$.
Here $C_v$ is the valley Chern number\cite{Niu10} of the gap (see Subsec. \ref{sect3d} below). 
The gap and valley Chern numbers in the vicinity of $\varphi = 0, 60, 120^{\circ}$
are $C_{g}=-2$ and $C_{v}=0$, while they are $C_{g}=0$ and $C_{v}=1$ ($C_{v}=-1$) 
in the region centered at $\varphi = 30^{\circ}$ ($\varphi =  90^{\circ}$). 
Figure 2 thus shows that the angle period is 120$^{\circ}$ rather than
60$^{\circ}$. Although the system has the same Chern number of 0 for $\varphi=30^{\circ}$ 
and $\varphi=90^{\circ}$, its valley Chern numbers at these $\varphi$ values have opposite signs.
Clearly, such a topological gap map provides us useful information for, e.g., designing photonic 
topological insulators and hence reflection-free one-way waveguides as well as for engineering 
topological phase transitions. Characteristics of the band gaps for the $\varphi=0^{\circ}$ 
and $\varphi=30^{\circ}$ cases are listed in Table I. 

\begin{figure*}
\includegraphics[width=16cm]{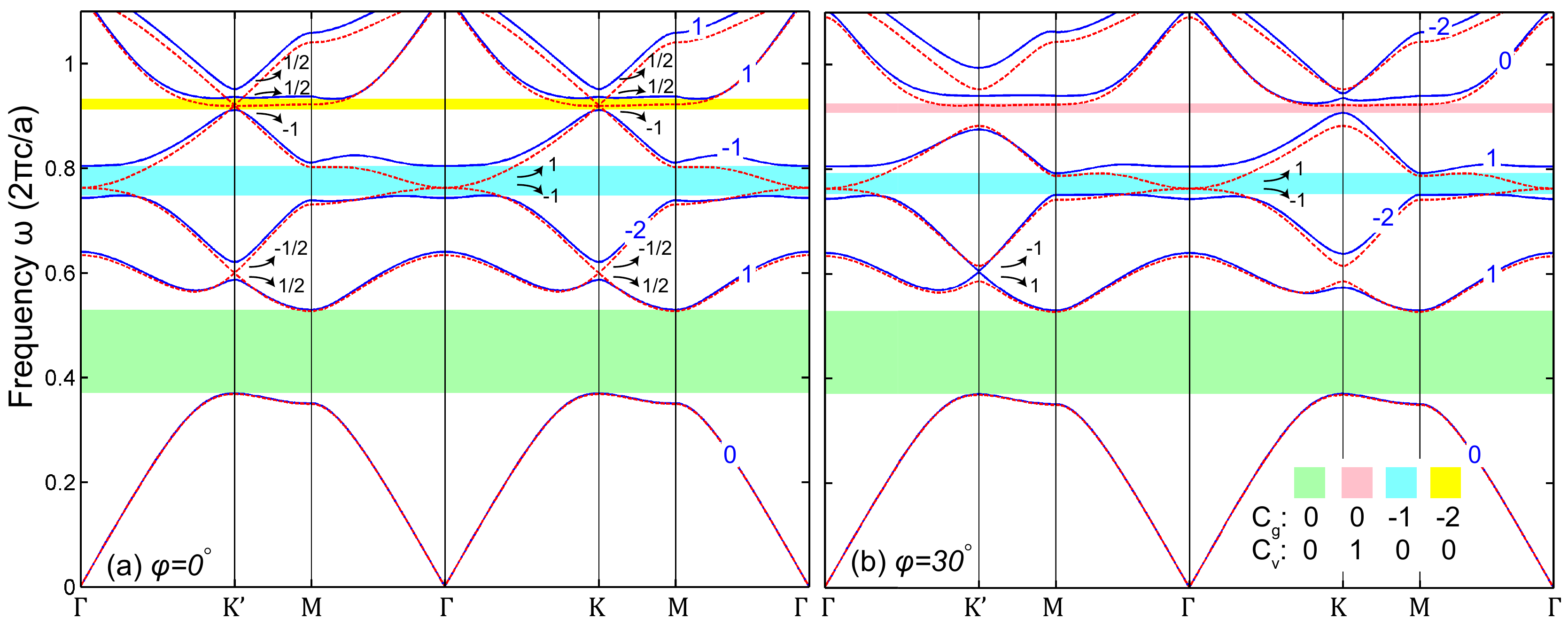}
\caption{\label{bands}
Band structures of the photonic crystal with (a) $\varphi=0^{\circ}$ and
(b) $\varphi=30^{\circ}$. Both $\mathcal{T}$-invariant ($\kappa=0$, red dotted lines)
and $\mathcal{T}$-symmetry broken ($\kappa=0.4$, blue solid lines) band structures are displayed.
In the $\kappa=0.4$ case, the band Chern numbers are labeled by integers.
The black arrows and integers illustrate how Chern number exchange between two adjacent bands.
Note that the second band gap (blue region) is a QAH phase with the gap Chern number $C_{g}=-1$. 
Interestingly, the third band gap in (a) is a QAH phase with $C_{g}=-2$ (yellow region),
while in (b) is a QVH phase with the valley Chern number $C_{v}= 1$ (pink region). 
}
\end{figure*}

\subsection{Bulk band structures}\label{sect3b}
To understand the formation mechanisms of the topological gaps and also other interesting properties of
the triangular photonic crystal, we present the calculated band structures for the  
$\varphi=0^{\circ}$ and $\varphi=30^{\circ}$ cases in Fig. \ref{bands}(a) and Fig. \ref{bands}(b), respectively.
The TM mode (i.e., $\mathbf{E}=E_z\hat{z}$) 
is considered here. To see how the band structure evolves when dielectric triangular rods
are replaced by gyromagnetic rods, both band structures of $\kappa=0.4$ and $\kappa=0$ are presented. 
Adding a non-zero off-diagonal element ($\kappa$) to the permeability tensor [Eq. (1)] breaks $\mathcal{T}$ symmetry.

Several observations can be drawn from an examination of Fig. \ref{bands}. Firstly, when $\kappa=0$ and
hence the system has the $\mathcal{T}$ symmetry, the energy bands (red dotted lines) 
along the $\Gamma$KM$\Gamma$ and $\Gamma$K$'$M$\Gamma$ lines are identical 
[see Fig. \ref{bands}(a) and Fig. \ref{bands}(b)]. This is because
the $\mathcal{T}$ symmetry ensures $\omega (-\mathbf{k})=\omega (\mathbf{k})$.\cite{joanBook} 
Secondly, when $\varphi=0^{\circ}$ and $\kappa=0$, there are doubly degenerate points 
(i.e., the Dirac points) at the $K$ and $K'$ points [see red dotted lines 
near $\omega = 0.6$ $2\pi c/a$ in Fig. \ref{bands}(a)]. 
In contrast, the $\varphi=30^{\circ}$ case has no such degenerate points 
[see red dotted lines in Fig. \ref{bands}(b)]. 
This results from the fact that the $\varphi=0^{\circ}$ geometry has 3 mirror $m_y$ 
reflections (i.e., 3 vertical reflection 
planes along $KK'$ direction) while the $\varphi=30^{\circ}$ structure lacks such mirror $m_y$ symmetry. 
In other words, the Dirac points at $K$ and $K'$ points in the $\varphi=0^{\circ}$ structure
occur because of the presence of both $\mathcal{T}$ and $m_y$ symmetries. 
Thirdly, Figure 3 shows that when $\kappa=0$, there are doubly degenerate points (i.e., massive
Dirac nodes) (see red dotted lines
close to $\omega = 0.75$ $2\pi c/a$) at the center of the BZ (the $\Gamma$ point). 
When the $\kappa$ becomes nonzero and hence $\mathcal{T}$ symmetry is broken, 
the double degeneracies are lifted and the Dirac nodes become gapped. 
When these Dirac nodes become gapped due to broken $\mathcal{T}$ symmetry,
the two bands would exchange $2\pi$ Berry phase (see Fig. 4), giving rise to a nontrivial gap.
Finally, Fig \ref{bands} (a) shows that there are triply degenerate nodal points
at the $K$ and $K'$ points near $\omega = 0.92$ $2\pi c/a$.
These rare triply degenerate nodal points are caused by the accidental degeneracy of a doubly
degenerate Dirac point and a nondegenerate band.
Fascinatingly, it was shown that photonic crystals with a triply degenerate nodal point at the
$\Gamma$ point may be used to realize zero-refractive-index metamaterials.\cite{Huang11} 
These interesting three-fold nodal points were recently found in topological phonic crystals.\cite{Lu15}
More recently, they were also predicted in
several electronic topological metals\cite{Weng16,Zhu16} and subsequentlty observed 
in topological semimetal MoP\cite{Lv17}. All these indicate
that the triply degenerate nodal points are attracting increasing attention 
in the field of electronic, phononic and photonic topological materials. 

The topological nature of a band gap in a broken $\mathcal{T}$ symmetry structure
can be characterized by the gap Chern number ($C_{g}$), 
as mentioned before in Sec. II.  
The calculated band Chern numbers displayed in Fig. 3 indicate that
the first band gap with $C_{g} = 0$ is topologically trivial in both $\varphi=0^{\circ}$ 
and $\varphi=30^{\circ}$ cases. In contrast, the second band gap is a QAH phase with $C_{g} = -1$ 
[see Fig. \ref{bands}(a) and Fig. \ref{bands}(b)]. 
Note that this band gap results from the lifting of the doubly degenerate Dirac nodes 
at the $\Gamma$ point by breaking the $\mathcal{T}$ symmetry, which exist in 
both $\varphi=0^{\circ}$ and $\varphi=30^{\circ}$ cases.
Remarkably, in the $\varphi=0^{\circ}$ case the third band gap is a high Chern number 
QAH phase with $C_{g} = -2$ [Fig. \ref{bands}(a)]. 
This interesting band gap results from gapping the rare triply degenerate
nodal points at the $K$ and $K'$ points by broken $\mathcal{T}$ symmetry ($\kappa = 0.4$). 
Thus, the gap Chern number ($|C_{g}|$) is larger than 1. 
Interestingly, this large Chern number gap would allow us to create multi-mode one-way edge states
and hence waveguides.\cite{skir14} In the $\varphi=30^{\circ}$ case, on the other hand, 
the band gap is a QVH phase with valley Chern number $C_{v} = 1$ [Fig. \ref{bands}(b)],
which will be discussed in the next Subsec. 
Note that the third gap already occurs when $\kappa=0$ and 
thus is not caused by broken $\mathcal{T}$ symmetry [Fig. \ref{bands}(b)].
Instead, it results from broken $m_y$ mirror symmetry.
Nonetheless, when $\kappa$ becomes nonzero (e.g., $\kappa = 0.4$), 
this gap at the $K$ point becomes larger. Interestingly, in contrast, this gap at the $K'$ point 
becomes smaller when $\kappa$ becomes nonzero.

\begin{figure*}
\includegraphics[width=16cm]{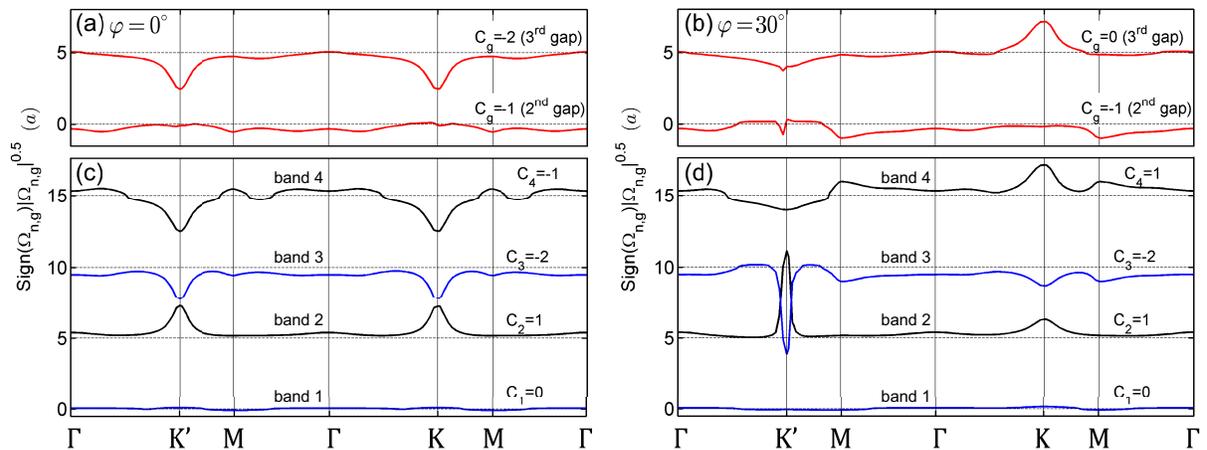}
\caption{\label{alongBerry}
Gap Berry curvatures $\Omega_{g}$ [(a) and (b)] and band Berry curvatures $\Omega_{n}$ [(c) and (d)], respectively,
for the photonic crystal with $\varphi=0^{\circ}$ [(a) and (c)] and $\varphi=30^{\circ}$ [(b) and (d)]. 
For clarity, sign($\Omega_{g}$)$|\Omega_{g}|^{1/2}$ for the third gap in (a) and (b)
has been shifted upwards by 5 $a$, and sign($\Omega_{n}$)$|\Omega_{n}|^{1/2}$ in (c) and (d) has been shifted upwards by $5a, 10a$ and $15 a$
for bands $n = 2, 3$ and $4$, respectively. The black dotted lines indicate the zero value positions.
}
\end{figure*}

Lastly, Fig. 3 shows that when $\kappa$ becomes nonzero, there is no direct overlap between
bands 2 and 3 in the entire BZ. Note that there is a small gap between bands 2 and 3 at $K'$
in Fig. 3(b). In other words, there is a continuous gap separating bands 2 and 3 throughout
the BZ and hence bands 2 and 3 have a well-defined band Chern number. Furthermore,
this gap is topologically nontrivial since it has a nonzero gap Chern number of $+1$.
Nonetheless, this gap is not marked on the gap map in Fig. 2 simply because it is
not a complete gap due to the indirect overlap between bands 2 and 3.

\subsection{Berry curvatures}\label{sect3c}
\begin{figure}
\includegraphics[width=6cm]{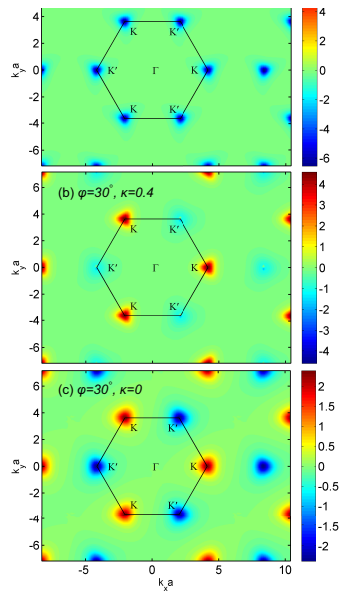}
\caption{\label{gapBerry}
Contour plots of the third gap Berry curvature $\Omega_{g}(\mathbf{k})$ of the triangular photonic
crystal with (a) $\varphi=0^{\circ}$ and $\kappa=0.4$ ($C_{g}=-2$), (b) $\varphi=30^{\circ}$ 
and $\kappa=0.4$ $(C_{g}=0)$,
and (c) $\varphi=30^{\circ}$ and $\kappa=0$ $(C_{g}=0)$.
The values on the color bars are in units of $a^2$, where $a$ is the lattice constant.
}
\end{figure}

In order to unravel how the $C_n$ numbers exchange between neighboring bands as well as the resultant $C_{g}$ arises, 
we calculate the band Berry curvatures $\Omega_n(\mathbf{k})$ for all the four bands
below the third band gap (see Fig. \ref{bands}) and plot these Berry curvatures for $\varphi=0^{\circ}$
and $\varphi=30^{\circ}$ in Fig. \ref{alongBerry}. Note that in Fig. 4 we plot sign($\Omega_{n,g})|\Omega_{n,g}|^{1/2}$ 
in order to better reveal the variation of the Berry curvatue with $\mathbf{k}$. Here we follow Ref. [\onlinecite{Guo08}] 
where the band-decomposed spin Berry curvatures helped to reveal the origin of the gigantic 
spin Hall conductivity in platinum metal. It turns out that gapping the degenerate points at $K$ and $\Gamma$  
cause quite different Berry curvatures.
We also evaluate the gap Berry curvatures $\Omega_{g}(\mathbf{k})$ given by 
\begin{equation} 
\label{eq5} 
\Omega_{g}(\mathbf{k})=\sum_n \Omega_n(\mathbf{k}) 
\end{equation}
where the summation is over all the band Berry curvature $\Omega_n(\mathbf{k})$ 
below the gap. An integration of $\Omega_{g}(\mathbf{k})$ over the BZ gives rise
to the gap Chern number $C_{g}$.\cite{skir14} The calculated $\Omega_{g}(\mathbf{k})$ for the
second and third gaps for $\varphi=0^{\circ}$
and $\varphi=30^{\circ}$ are displayed as a function of $\bf{k}$ 
in Fig. \ref{alongBerry}. The contour plots of $\Omega_{g}(\mathbf{k})$ on the 
$k_z =0$ plane for the third gap for $\varphi=0^{\circ}$
and $\varphi=30^{\circ}$ are given in Fig. \ref{gapBerry}. 

Figure \ref{alongBerry}(a) shows that for bands 2-4, all the band Berry curvatures $\Omega_n$
peak at $K$ and $K'$. Furthermore, the signs of the peaks for bands 2 and 3 
are opposite. As mentioned before, with the $\mathcal{T}$ symmetry, the $2$nd and $3$rd 
bands touch at $K$ and $K'$ and thus form the massless Dirac points [Fig. 3(a)].
However, when the $\mathcal{T}$ symmetry is broken, these Dirac points become gapped and 
the $2$nd and $3$rd bands exchange $\pm 2\pi$ Berry phase [Berry phase 
$\phi_n =2\pi C_n$ given by Eq. \eqref{eq2}] and hence $C_n=\pm 1$. This gives rise 
to the peaks with opposite signs in $\Omega_n$ at these Dirac points.
Interestingly, if both bands are below the gap of interest, their contributions to
the gap Chern number cancel each other and thus become diminished.
Therefore, in the present case, the main contribution to the gap 
Berry curvature and hence the gap Chern number of the third gap comes mainly 
from the 4th band which has the pronounced peaks at $K$ and $K'$ [Fig. 4(a)].
Interestingly, in the case of $\varphi=30^{\circ}$, bands 2 and 3 have a large peak at $K'$ 
but a small peak at $K$. Nevertheless, these two peaks with opposite signs cancel each other
and hence the dominant contribution to the gap Berry curvature comes predominantly from band 4. 
However, the band Berry curvatures of band 4 in the regions centered at $K$ and $K'$  
have opposite signs [see Fig. \ref{alongBerry}(b) and Fig. 5(b)] and consequently, 
the gap Chern number of the third gap is zero.
In contrast, Fig. \ref{alongBerry} shows that the $\Omega_{g}$ of the second gap ($C_{g}=-1$)
varies more smoothly over the BZ, and this could be attributed to the much larger
gap openned at the massive Dirac point on the $\Gamma$ point. Finally, this $\Omega_{g}$
has a small peak at $M$ instead (Fig. \ref{alongBerry}).

Figures 3 and 4 can also help us to better understand how $C_n$ exchange at the degenerate points
when the $\mathcal{T}$ symmetry is broken. For instance, Fig. 4(c) shows that both $\Omega_2$ and $\Omega_3$ 
have a peak at the $K$ and $K'$, but with opposite signs. This suggests that band 2 and band 3 
exchange $\pm 1/2$ Chern number at $K'$ and $K$, as illustrated by the black arrows at $K'$ and $K$ in Fig. 3(a). 
Likewise, $\Omega_3$ and $\Omega_4$ have similar behaviors near the $\Gamma$. Here the exchanged Chern numbers
have opposite signs and this is because gapping the Dirac point at $\Gamma$ would result in an exchange 
of $\pm 1$ Chern numbers. Indeed, $C_3=-1-1 = -2$ in which band 3 receive $-1$ from both band 2 and band 4. 
Interestingly, we find that after gapping the triple degeneracy, the 4th band receives $-1$ Chern number 
from the 5th and 6th bands at both $K'$ and $K$ [see Fig. 3(a)] such  
that $C_4=1-1-1=-1$ where $+1$ Chern number comes from band 3, 
being consistent with the negative peaks seen at the $K$ and $K'$ in Fig. 4(a).

Finally, let us examine the calculated gap Berry curvature $\Omega_{g}(\mathbf{k})$
over the entire 2D BZ. Figure \ref{gapBerry}(a) shows clearly that in the $\varphi=0^{\circ}$ case, 
the $\Omega_{g}(\mathbf{k})$ of the third gap has six pronounced negative peaks, respectively, at 
six $K$ and $K'$ points, although it is rather flat with a small negative value 
of $|\Omega_{g}(\mathbf{k})| < 0.5 a^2$ over the rest of the BZ. 
As mentioned above, these six peaks are caused by the lifting of the triply degenerate nodal points 
at both $K$ and $K'$ points [see Fig. \ref{bands}(a)] due to the replacement of the dielectric rods 
($\kappa=0$) by the gyromagnetic rods ($\kappa=0.4$) which breaks the $\mathcal{T}$ symmetry. 
These six peaks are identical because the $\Omega_{g}(\mathbf{k})$ distribution has the $C_{6v}$ symmetry.  
Therefore, the integration of the $\Omega_{g}(\mathbf{k})$ gives rise to a nonzero 
gap Chern number $C_{g}$ of -2. Thus, the 3rd gap is a high Chern number QAH phase.

Interestingly, Figs. \ref{gapBerry}(b) and \ref{gapBerry}(c) reveal that the three negative peaks 
at the three $K$ points in the $\varphi=0^{\circ}$ case become three large positive peaks 
in the $\varphi=30^{\circ}$ case, caused by a simple rotation of angle $\varphi$ from $0^{\circ}$
to $30^{\circ}$. The $\Omega_{g}(\mathbf{k})$ in the rest of
the BZ is almost zero. The $\Omega_{g}(\mathbf{k})$ distribution now has the $C_{3v}$ symmetry. 
Consequently, the contribution from the three positive peaks at the three $K$ points to the $C_{g}$ 
cancels that from the three negative peaks at the $K'$ points, resulting in $C_{g}=0$. 
Interestingly, this unique $\Omega_{g}(\mathbf{k})$ pattern is identical to that in the electronic
band structure of MoS$_2$ monolayer\cite{Feng12,Mak14}, and thus is the signature of
the QVH phase with nonzero valley Chern number of $C_v = +1$.\cite{Ma16,Chen17}
Here $C_v$ is defined as the difference between two valley indices $C^{K}$ and $C^{K'}$, 
i.e., $C_v = C^{K} - C^{K'}$.\cite{Ma16,Chen17} The valley index $C^{K}$ ($C^{K'}$)
can be obtained by integrating the gap Berry curvature $\Omega_{g}(\mathbf{k})$ over
half of the IBZW in the vicinity of the $K$ ($K'$) point. In the $\varphi=0^{\circ}$ case,
clearly, this would give rise to $C^{K} = +1/2$ and $C^{K'} = -1/2$ [see Fig. \ref{gapBerry}(c)], 
thus leading to $C_v = +1$. 
A further rotation of $\varphi$ by another $30^{\circ}$ would make the system return to the case 
of $\varphi=0^{\circ}$, i.e., the cases of $\varphi=60^{\circ}$ and $\varphi=0^{\circ}$ have 
the same $\Omega_{g}(\mathbf{k})$ distribution [Thus, the $\Omega_{g}(\mathbf{k})$ of $\varphi=60^{\circ}$ 
is not shown here]. Another rotation of $\varphi$ by $30^{\circ}$ to $\varphi=90^{\circ}$  
would lead to a $\Omega_{g}(\mathbf{k})$ pattern that is identical to that of the $\varphi=30^{\circ}$ case
except a swap of $K$ and $K'$ points [Thus, the $\Omega_{g}(\mathbf{k})$ of $\varphi=90^{\circ}$ 
is not shown here]. However, the two valley indices would swap signs, thereby resulting in
a valley Chern number of  $C_v = -1$. This shows that the properties of the system as a function of $\varphi$ 
has a period of $120^{\circ}$.

\subsection{Chiral edge states}\label{sect3d}
The principle of bulk-edge correspondence guarantees that gapless one-way (chiral) edge states occur at
the interface between two topologically different bulk insulators and the number of 
these edge states equals to the difference in the gap Chern numbers of the two insulators. 
(see, e.g., [\onlinecite{joan14}]).  To further study these fascinating edge states,
we calculate the edge band diagram, i.e., frequency dispersions $\omega (\mathbf{k_{\parallel}})$ 
for the wave vector ${k_{\parallel}}$ along the edge of the photonic crystal for both $\varphi=0^{\circ}$ 
and $\varphi=30^{\circ}$, as displayed in Figs. 6(a) and 6(b), respectively.
In this calculation, a supercell consisting of one unit cell along the interfacial $x$-direction 
[also the propagation direction (${k_{\parallel}}=\hat{x}k_x$)] and 20 unit cells of $\varphi=0^{\circ}$ [Fig. 6(a)] 
and $\varphi=30^{\circ}$ [Fig. 6(a)] photonic crystals along $KK'$ direction 
with both ends terminated by a perfect electric conductor (PEC). Here the PEC is adopted to 
mimic a metal in the microwave region.\cite{joan08,skir14} 

\begin{figure}
\includegraphics[width=8cm]{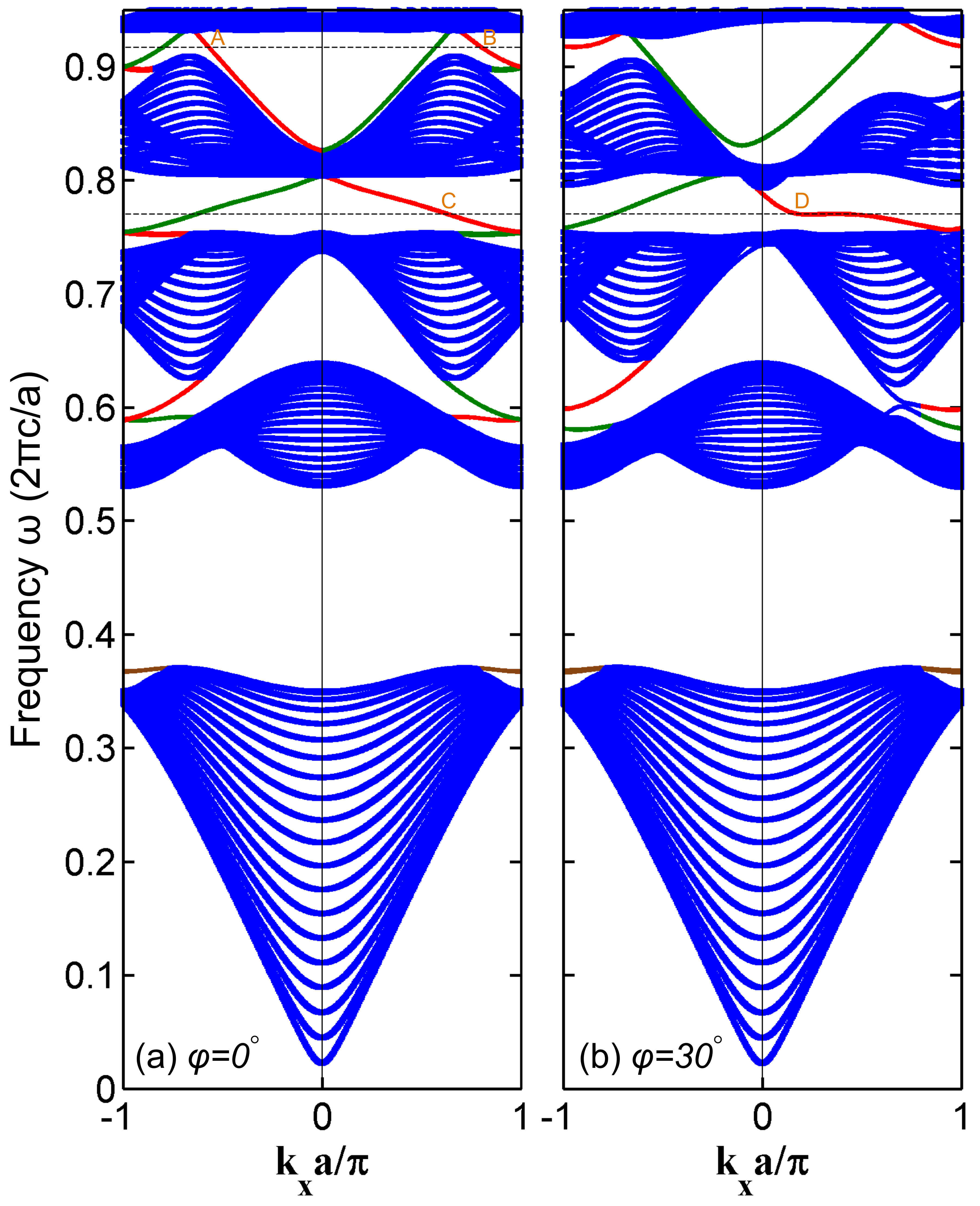}
\caption{\label{edge}
Edge band diagram of the gyromagnetic photonic crystal ribben of (a) $\varphi=0^{\circ}$ 
and (b) $\varphi=30^{\circ}$ with both edges terminated by a perfect electric conductor. 
The red (green) lines represent upper (lower) edge bands, while the blue curves denote bulk projected bands. 
The brown lines at the top of bulk band 1 denote doubly degenerate upper and lower edge bands.
The field profiles of the edge states at the cross points of the dotted horizontal lines 
and edge bands and marked as $A, B, C, D$, are shown in Fig. 7.
The slopes of gapless edge bands (i.e., their group velocities $\partial \omega / \partial k $)
indicate the wave propagation directions. One edge band for each edge is observed in the second gap
while (a) two and (b) zero edge bands at each edge exist in the third gap. These agree well with
the calculated $C_g$ shown in Fig. 3.
}
\end{figure}

\begin{figure}
\includegraphics[width=8cm]{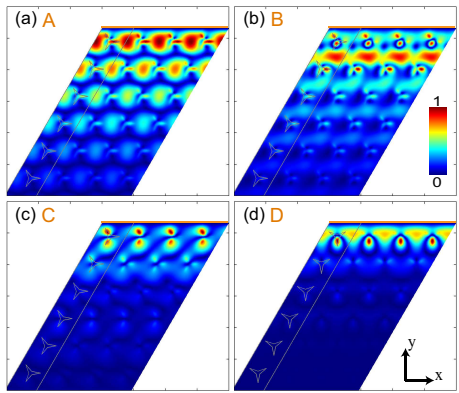}
\caption{\label{edgefield}
Electric field profiles ($|E_z|$) of the edge states labelled $A$, $B$, $C$ and $D$ in Fig. \ref{edge}.
The color bar is in units of V/m. 
In each panel, the orange line on the top represents the metallic wall at the edge of the metastructure.}
\end{figure}

Figure \ref{edge} shows that apart from bulk projected bands (blue curves), there are one gapless 
edge band (red lines) in the 2nd gap region, which agree well with calculated $C_g=-1$ shown in Fig. 3. 
Similarly, in Fig. \ref{edge}(a) we can see two edge bands in the 3rd gap ($C_g=-2$). 
Note that the slopes of the edge bands indicate the direction of propagation (i.e., their group 
velocities $\partial \omega / \partial k $). In Fig. \ref{edge}, the negative slopes of edge bands 
suggest that the waves would propagate towards the $-\hat{x}$ direction. On the other hand, 
as shown in Fig. \ref{edge}(b), no gapless one-way edge band is observed in the 3rd gap, being consistent
with $C_g=0$ in Fig. 3. All these observations show that 
bulk-edge correspondence is satisfied. Moreover, the $|(E_z)|$ field profiles of the edge states 
labelled as $A$, $B$, $C$ and $D$ are shown in Fig. \ref{edgefield}. Clearly, the electric fields 
of edge states $C$ and $D$ from the 2nd band gap are more confined to the edge 
than that of the $A$ and $B$ states in the 3rd gap (Fig. 7), simply because the 2nd band gap is wider
(Table I). Finally, we can also see one chiral edge band in the incomplete gap near 0.6 $2\pi c/a$ in Fig. 6,
as can be expected from its nonzero gap Chern number of $+1$ shown in Fig. 3. 

\subsection{Reflection immune one-way waveguides}\label{sect3e}
\begin{figure}
\includegraphics[width=8cm]{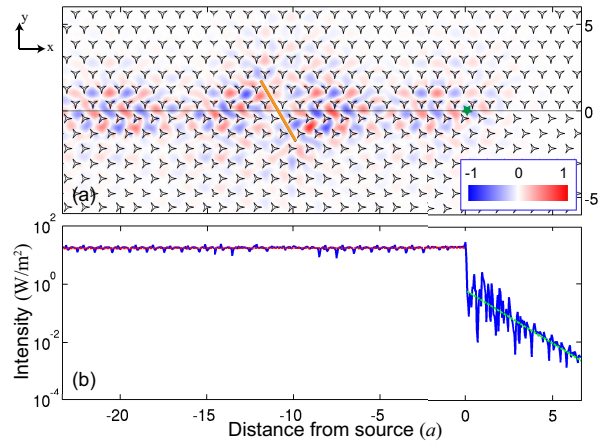}
\caption{\label{topoWG}
Numerical simulation of wave propagation in an interfacial state
between two topologically distinct photonic crystals of
$\varphi=0^{\circ}$ ($C_{g}=-2$) and $\varphi=30^{\circ}$ ($C_{g}=0$) with
a point source  of frequency $\omega = 0.92$ $2 \pi c/a$ within the third topological gap (see Fig. 3).
(a) Calculated Re$(E_z)$ distribution. The values on the color bar are in units of V/m.
The upper region is the $\varphi=30^{\circ}$ photonic crystal and the lower region is the
 $\varphi=0^{\circ}$ photonic crystal. The green star denotes the point source.
The orange line at -11$a$ represents a metallic plate of thickness $0.1a$ and width $4a$
with the plate surface tilted $\varphi=30^{\circ}$ away from $y$-axis.
(b) Intensity ($I$) as a function of the distance from the point source.
Here $I$ is defined as the integrated intensity over the cross-section divided 
by the area of the cross-section.
}
\end{figure}

In this subsection, we should demonstrate that topologically protected one-way edge modes propagate at the 
interface of a Chern photonic crystal 
with another topologically different material by performing numerical simulations 
of electromagnetic (EM) wave propagations along the interface. 
We will also access their application potentials such as reflection free one-way waveguides.
Both the second and third topological gaps of  the $\varphi=0^{\circ}$ photonic crystal [see Fig. 3(a)] 
will be considered.  Let us first study the interface between two topological distinct photonic crystals,
namely, $\varphi=0^{\circ}$ ($C_g=-2$) and $\varphi=30^{\circ}$ ($C_g=0$), with the source frequency 
in the third band gap.  As illustrated in Fig. \ref{topoWG}(a), the simulation 
system used consists of the $\varphi=0^{\circ}$ crystal (lower region)
and the $\varphi=30^{\circ}$ crystal ($C_{g}=0$) (upper region). A point source, which radiates in all directions
with frequency $\omega = 0.92$ $2\pi c/a$ in the third gap, is placed at the origin (green star) in the interface. 
Moreover, as an obstacle to the EM wave propagation, a metallic plate (orange line) of thickness $0.1a$ 
and width $4a$ is inserted at -11$a$ 
with the plate surface tilted $\varphi=30^{\circ}$ away from $y$-axis [Fig. \ref{topoWG}(a)].
The Re$(E_z)$ field distribution displayed in Fig. \ref{topoWG}(a) demonstrates firstly that the EM 
wave cannot enter the upper and lower regions because the operating frequency is in the band gap, 
and secondly that it cannot go right (the $\hat{x}$ direction) because it is forbidden by the 
topological nature of the edge state. For example, the ratio ($I_-$/$I_+$) of the light intensity 
at -17$a$ to that at 6$a$ is $\sim 3.8\times10^3$.
Such significant difference is the signature of unidirectionality.
Therefore, the EM wave effectively can only propagate toward the $-\hat{x}$ direction.
Furthermore, as expected, the EM wave can circumvent the metallic plate and continue traveling 
towards the $-\hat{x}$  without loss
[see the calculated intensity as a function of the distance from the source in Fig. \ref{topoWG}(b)].
In particular, the ratio of the transmitted intensities at -17$a$ to that at -7$a$ is $\sim$0.990,
indicating that the scattering due to the obstacle (the metal plate) is minimal.

\begin{figure}
\includegraphics[width=8cm]{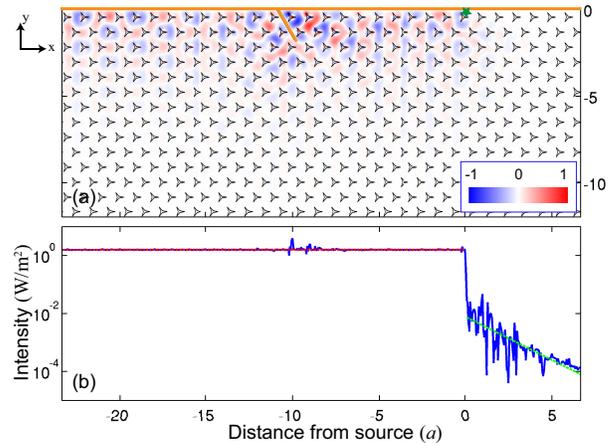}
\caption{\label{PEC3gap}
Numerical simulation of wave propagation in an interfacial state
between the photonic crystal of $\varphi=0^{\circ}$ ($C_{g}=-2$) and a metallic wall with
a point source  of frequency $\omega = 0.92$ $2 \pi c/a$ within the third gap [see Fig. 3(a)].
(a) Calculated Re$(E_z)$ distribution. The values on the color bar are in units of V/m.
The lower region is the $\varphi=0^{\circ}$ photonic crystal and the metallic wall (orange line)
is on the top of the photonic crystal. The green star denotes the point source.
The orange line at -11$a$ represents a metallic plate of thickness $0.1a$ and width $2a$
with the plate surface tilted $\varphi=30^{\circ}$ away from $y$-axis.
(b) Intensity ($I$) as a function of the distance from the point source.
Here $I$ is defined as the integrated intensity over the cross-section divided
by the area of the cross-section.}
\end{figure}

\begin{figure}
\includegraphics[width=8cm]{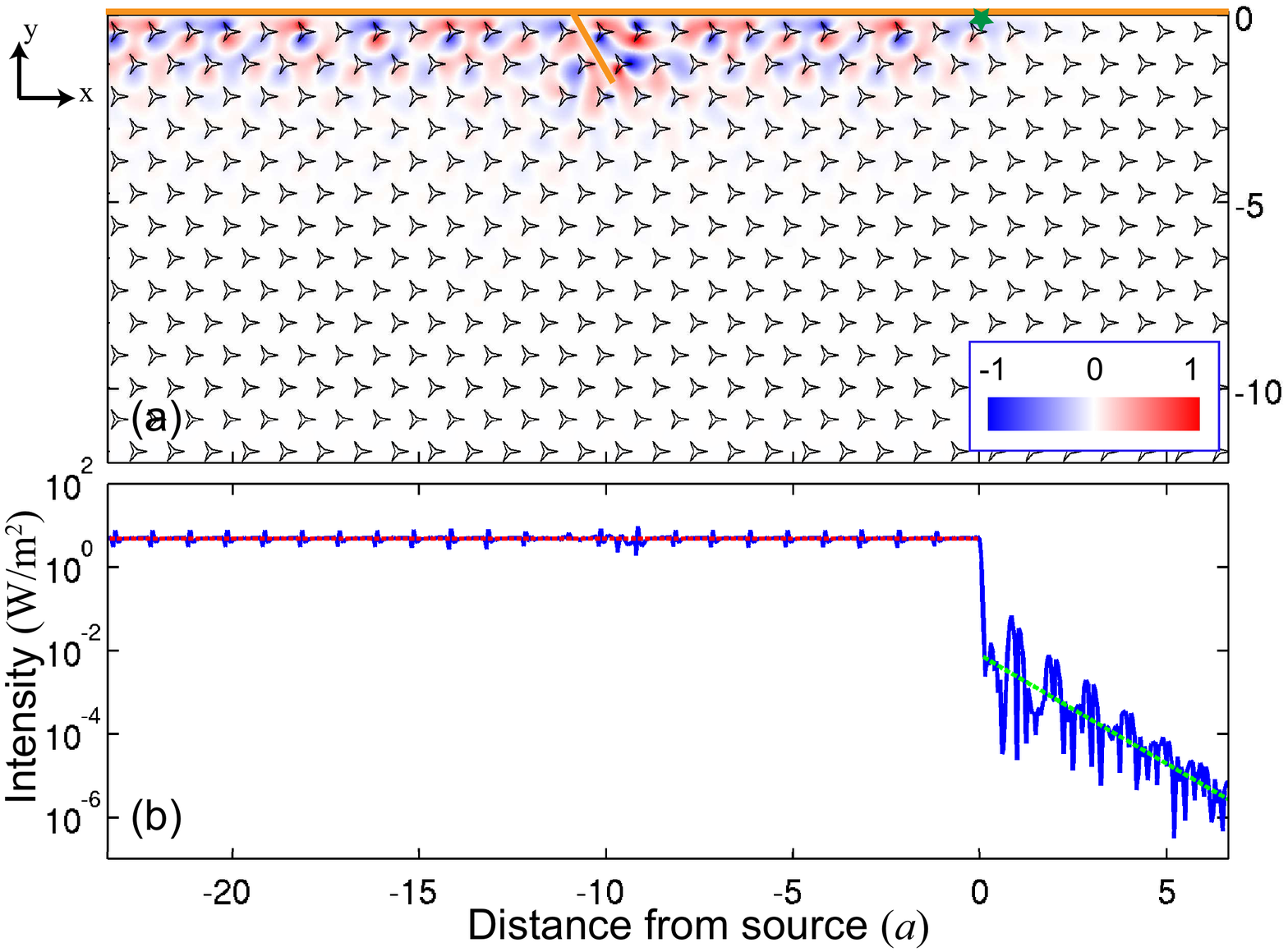}
\caption{\label{PEC2gap}
Numerical simulation of wave propagation in an interfacial state
between the photonic crystal of $\varphi=0^{\circ}$ ($C_{g}=-1$) and a metallic wall with
a point source  of frequency $\omega = 0.78$ $2 \pi c/a$ within the second gap [see Fig. 3(a)].
(a) Calculated Re$(E_z)$ distribution. The values on the color bar are in units of V/m.
The lower region is the $\varphi=0^{\circ}$ photonic crystal and the metallic wall (orange line)
is on the top of the photonic crystal. The green star denotes the point source.
The orange line at -11$a$ represents a metallic plate of thickness $0.1a$ and width $2a$
with the plate surface tilted $\varphi=30^{\circ}$ away from $y$-axis.
(b) Intensity ($I$) as a function of the distance from the point source.
Here $I$ is defined as the integrated intensity over the cross-section divided
by the area of the cross-section.}
\end{figure}

For the application as a waveguide, the surface of a photonic topolical insulator 
is usually covered with a metal (e.g., copper) wall. Therefore, we also perform similar 
simulations for the interface between the $\varphi=0^{\circ}$ photonic crystal and 
a metallic wall which is topologically trivial. In principle, the larger the topological gap, 
the better the localization of the EM wave in the interfacial region.
As a result, from the viewpoint of the application of the edge modes as one-way waveguides, 
it would be advantegeous to use a larger band gap. Therefore, we consider 
the $\varphi=0^{\circ}$ photonic crystal with the source frequency 
being $\omega = 0.92$ $2\pi c/a$ within the third gap ($C_{g}=-2$, $E_g = 2.3$ $\%$) and 
also $\omega = 0.78$ $2\pi c/a$ within the second gap ($C_{g}=-1$, $E_g = 7.3$ $\%$). 
The calculated Re$(E_z)$ field and also the transmission intensity are displayed in 
Fig. \ref{PEC3gap} and Fig. \ref{PEC2gap}, respectively. Indeed, Fig. \ref{PEC2gap}(a) 
shows that the Re$(E_z)$ field is more confined to the interfacial region compared 
to that shown in Fig. \ref{PEC3gap}. Moreover, the unidirectionality of the
edge mode in the second gap is much better than that of the third gap. This is reflected in the fact that
the ratio ($I_{-}/I_{+}$) of the wave intensity at -17$a$ to that at 6$a$ in Fig. \ref{PEC2gap}
is nearly three orders of magnitude higher than in Fig. \ref{PEC3gap} (see Table I). 
Interestingly, Fig. 2 shows that the topological nature of the second gap does not depend 
on the rotation angle $\varphi$. As a result, the reflection-free one-way waveguiding would be 
very robust against the disorders such as the imperfect alignments of the triangular rods 
in the photonic crystal introduced during the waveguide fabrication processes.

The main properties of the band gaps such as the gap sizes, the gap Chern numbers ($C_g$), 
the numbers of edge states ($N_g$) and on-and-off edge intensity ratio ($I_{-}/I_{+}$)
are listed in Table. I. Firstly, Table I shows that the second gap sizes ($C_g=-1$) 
in both $\varphi=0^{\circ}$ and $\varphi=30^{\circ}$ cases are large, being comparable 
to some well-known designs using circular rods.\cite{joan08,skir14} 
Secondly, that $|C_g|=N_g$ comfirms the bulk-edge correspondence, as mentioned before. 
Lastly, the large magnitudes of $I_{-}/I_{+}$ especially of 
$\sim 10^{6}$ for the second gap indicates that the waveguides
made of the proposed photonic crystals would have a very high unidirectionality.
This unidirectionality also manifests itself as a sharp drop in the intensity 
at the right-hand side of the point source towards the $+\hat{x}$ direction in Figs. 8-10 (b).

\begin{figure}
\includegraphics[width=6cm]{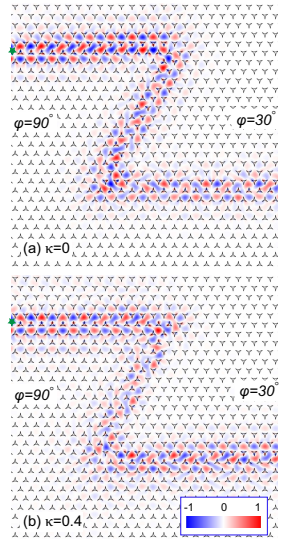}
\caption{\label{Zshape}
Numerical simulation of wave propagation in a valley interfacial state along a Z-shape 
bend between two valley-contrasting photonic crystals of $\varphi=30^{\circ}$ and $\varphi=90^{\circ}$.
(a) $\kappa=0.0$ and (b) $\kappa=0.4$. 
The green star denotes the point source operating at (a) $\omega = 0.9$ and 
(b) $0.91$ $2 \pi c/a$. The values on the color bar denote the Re$(E_z)$ in units of V/m.
}
\end{figure}

\subsection{Valley Hall edge states and light propagation in a Z-shape bend}\label{sect3f}
The above examination of the $\Omega_{g}(\mathbf{k})$ of the third gap over the BZ in Sec. \ref{sect3c} 
reveals the interesting QVH phase in both $\varphi=30^{\circ}$ (valley Chern number $C_v = +1$) 
and $\varphi=90^{\circ}$ (valley Chern number $C_v = -1$) cases. 
The occurrence of this QVH phase is caused by a rotation of the triangular rods by
$30^{\circ}$ from $\varphi=0^{\circ}$ to $\varphi=30^{\circ}$. 
In the $\varphi=0^{\circ}$ case where both the $\mathcal{T}$ symmetry and the mirror symmetry $m_y$ exist, 
there are six Dirac cones of bands $4-6$ (Fig. \ref{bands}(a), red curves) at six $K$ and $K'$ points, respectively.
When the system is transformed to the $\varphi=30^{\circ}$ case, the mirror symmetry $m_y$ 
is broken and the Dirac points become gapped (Fig. \ref{bands}(b), red curves), leading to
nonzero Berry curvature with the odd symmetry,\cite{Niu10} i.e., $\Omega_{g}(-\mathbf{k})=-\Omega_{g}(\mathbf{k})$
[see also Fig. 5(c)].
An integration of $\Omega_{g}(\mathbf{k})$ over the BZ is zero, as required by the $\mathcal{T}$ symmetry. 
As mentioned before, the $\varphi=90^{\circ}$ case has the same $\Omega_{g}(\mathbf{k})$ pattern 
as that of the $\varphi=30^{\circ}$ case except a swap of $K$ and $K'$ points, i.e., the third band gap
of  the $\varphi=90^{\circ}$ case is also a QVH phase but with $C_v = -1$.

Surprisingly, when $\mathcal{T}$ symmetry is broken (i.e., $\kappa=0.4$), the gap Berry phase 
of the third gap remains zero. The topology of bands $4-6$ remains unaltered
and this can also be seen by comparing bands $4-6$ in the vicinity of the $K$ and $K'$ points in Fig. \ref{bands}(b)
before (red curves) and after (blue curves) the introduction of the gyromagnetic material.
Therefore, the third gap in the $\kappa=0.4$ and $\varphi=30^{\circ}$ case is
also a QVH phase with $C_v = +1$. Since $\mathcal{T}$ symmetry is now broken, 
the QVH phase should be called the pseudo-QVH phase, as for the pseudo-quantum 
spin Hall phases\cite{Chen11,Hsu17}.
Nonetheless, the profiles of the Berry curvatures at $K$ and $K'$ valleys now become
different, as shown in Fig. \ref{gapBerry}(b).

It is known that the edge states in a QVH photonic topological insulator is valley-dependent and
thus the intervalley scattering along the zigzag boundary\cite{shvets13,Chen17} is suppressed.
This would lead to valley-protected robust wave propagation in a Z-shape bend between
two QVH photonic crystals with different valley Chern numbers. Therefore, to explicitly 
verify the QVH phases in the $\varphi=30^{\circ}$ and $\varphi=90^{\circ}$ crystals, 
we performed numerical simulations of light propagation
along a Z-shape interface between these two topologically distinct crystals, as shown in Fig. \ref{Zshape}. 
Figure \ref{Zshape}(a) shows that in the $\kappa=0.0$ case, the wave can turn around 
the two sharp corners, thus proving the existence of the QVH edge states at the interface.
The same situation also occurs in the $\kappa=0.4$ case [\ref{Zshape}(b)], thereby suggesting
that the QVH edge states also exist in the interface between the $\kappa=0.4$ photonic crystals. 
Therefore, the results of these numerical simulations indicate that the QVH photonic crystal waveguides
presented here are not the ordinary photonic crystal waveguides and could have promising applications 
for, e.g., designng photonic valleytronic devices\cite{Ma16}.

\section{conclusions}\label{sect4}
In conclusion, we have carried a comprehensive theoretical study on the topological phases 
in 2D photonic crystals without  $\mathcal{T}$ and $\mathcal{P}$ symmetries.
As an example, we consider a hexagonal lattice consisting of
triangular gyromagnetic rods. Here the gyromagnetic material breaks
$\mathcal{T}$ invarience while the triangular rods breaks $\mathcal{P}$ symmetry.
Remarkably, we discover that the photonic crystal houses QAH phases
with different gap Chern numbers ($C_g$) including $|C_g| = 2$ 
as well as QVH phases with contrasting valley Chern numbers ($C_v$).
Moreover, phase transitions among these topological phases, such as from QAH to QVH
and vice versa, can be realized by a simple rotation of the orientation of the rods.
Our band theoretical analyses reveal that the Dirac nodes at the $K$ and $K'$ valleys
in the momentum space are produced and protected by the mirror symmetry ($m_y$) instead
of the $\mathcal{P}$ symmetry in the $\mathcal{P}$-invariant crystals, and they 
become gapped when either $\mathcal{T}$ or $m_y$ symmetry is broken, leading to a QAH or QVH
phase, respectively. Furthermore, the high Chern number ($C_g = -2$) QAH phase
arises when the rare triply degenerate nodal points\cite{CTCh12,Lu15} rather than pairs
of Dirac nodes\cite{skir14} are gapped by breaking $\mathcal{T}$ symmetry.
Therefore, our proposed photonic crystal would provide a platform for exploring transitions
among intriguing topological phases which may be very difficult to realize in electronic systems.

Our electromagnetic simulations of wave propagation either along the edges of our crystal
capped with a metal wall or in the interfaces between two variants of
our crystal with different rod orientations, demonstrate reflection-immune one-way light
transports in both straight interfaces and Z-shape bends. Furthermore, we find
that the second topologically nontrivial gap is not only large but also independent of the
orientation of the triangular rods, and thus the topologically protected one-way
wave propagation would be very robust against disorders such as misalignments of the
triangular rods in the photonic crystal introduced during the fabrication processes.
Therefore, our proposed crystal would also have promising potentials for device 
applications in photonics such as reflection-free waveguides and topological one-way circuits. 

\begin{acknowledgments}
This work is supported by the Ministry of Science and Technology, the National Center for Theoretical Sciences, 
Academia Sinica and the Kenda Foundation of The R.O.C.
\end{acknowledgments}


\end{document}